\documentclass[10pt]{article}

\usepackage{amsmath}
\usepackage{amssymb}

\usepackage{graphicx}

\usepackage{cite}

\usepackage{color} 
\newcommand{\be}{\begin{eqnarray}}
\newcommand{\ee}{\end{eqnarray}}

\usepackage[labelfont=bf,labelsep=period,justification=raggedright]{caption}

\makeatletter
\renewcommand{\@biblabel}[1]{\quad#1.}
\makeatother

\date{}

\begin{document}

\begin{flushleft}
{\Large
\textbf{Folding Pathways of a Knotted Protein with a Realistic Atomistic Force Field}
}
\\
 Silvio a Beccara$^{1}$, Tatjana \v{S}krbi\'c$^{2}$, Roberto Covino $^{3,4}$, Cristian Micheletti $^{5}$, Pietro Faccioli $^{3,4~\ast}$
\\
\bf{1}, LISC, Bruno Kessler Foundation, Trento, Italy.
\\
\bf{2}  ECT*, Bruno Kessler Foundation, Trento, Italy.
\\
\bf{3} Physics Department, University of Trento, Trento, Italy.
\\
\bf{4} INFN, Gruppo Collegato di Trento, Trento, Italy.
\\
\bf{5} SISSA and CNR-IOM Democritos, Trieste, Italy.
\\
$\ast$ E-mail: Corresponding faccioli@science.unitn.it
\end{flushleft}

\section*{Abstract}
 We report on atomistic simulation of the folding of a
  natively-knotted protein, MJ0366, based on a realistic force field. To
  the best of our knowledge this is the first reported effort where a
  realistic force field is used to investigate the folding pathways of a
  protein with complex native topology. By using the dominant-reaction
  pathway scheme we collected about 30 successful folding trajectories
  for the 82-amino acid long trefoil-knotted protein.  Despite the
  dissimilarity of their initial unfolded configuration, these
  trajectories reach the natively-knotted state through a remarkably
  similar succession of steps. In particular it is found that knotting
  occurs essentially through a threading mechanism, involving the
  passage of the C-terminal through an open region created by the
  formation of the native $\beta$-sheet at an earlier stage. The
  dominance of the knotting by threading mechanism is not observed in
  MJ0366 folding simulations using simplified, native-centric
  models. This points to a previously underappreciated role of concerted
  amino acid interactions, including non-native ones, in aiding the
  appropriate order of contact formation to achieve knotting.
 
\section*{Author Summary}

It has been recently observed that the native structure of proteins can
contain knots. These are formed during the folding process and are
tightened in a specific (i.e. native) location, along the poly-peptide
chain. The existence of knots hence implies a high degree coordination
of local and global conformational changes, during the folding reaction.
In this work we investigate how the knot is formed and what are the
dynamical mechanisms which drive the self-entanglement process.  To this
end, we report on the first atomistically detailed numerical simulation
of the folding of a knotted protein, based on a realistic description of
the inter-atomic forces.  These simulations show that the knot is formed
by following a specific sequence of contacts. The comparison of the
findings with those based on simplified folding models suggest that the
productive succession of contacts is aided by a concerted interplay of
amino acid interactions, arguably including non-native ones.

\section*{Introduction}
Natively-knotted proteins are increasingly studied as a new paradigm of
``multiscale'' folding coordination, which leads to establishing the
native knot in the native position starting from the unknotted
newly-translated
state~\cite{yeates_review,Taylor:2007:Comput-Biol-Chem:17500039,Mallam:2009:FEBS-J:19077162,virnau_review}. Intuitively,
the pathways associated to this process appear so improbable and prone
to misfolding that it was long held that naturally occurring proteins
would be protected against the occurrence of knots.  This {\em a priori}
expectation, which has a sound statistical basis
\cite{Lua:2006:PLoS-Comput-Biol:16710448,Potestio:2010:PLoS-Comput-Biol:20686683},
was so strong radicated that only several years after the publication of
the human carbonic anhydrase II structure
\cite{Eriksson:1988:Proteins:3151019} it was realized that it actually
accommodated a knot~\cite{Mansfield:1994:Nat-Struct-Biol:7656045}. Since
then, hundreds of instances of naturally-occurring knotted proteins have
been found and they now account for about 2\% of the protein data bank
(PDB) entries\cite{Potestio:2010:PLoS-Comput-Biol:20686683}.

The salient aspects of the folding phenomenology of several knotted
proteins have been recently probed by various experiments (for recent
reviews see
refs.~\cite{yeates_review,Taylor:2007:Comput-Biol-Chem:17500039,Mallam:2009:FEBS-J:19077162,virnau_review}). These
studies have demonstrated that newly translated, unknotted proteins, can
fold into the native knotted structure without the assistance of
chaperones
\cite{King:2010:Proc-Natl-Acad-Sci-U-S-A:21068371,Mallam:2012:Nat-Chem-Biol:22179065},
though the latter can significantly speed up the
process~\cite{Mallam:2012:Nat-Chem-Biol:22179065}.
The details of the concerted backbone movements that lead to the
self-tying of the protein in the native knot remain, however, beyond
reach of current experimental techniques.  In this regard, numerical
investigations can aptly complement experimental ones, by providing
valuable insight into the repertoire of viable modes of knot formation,
the stage at which the knot is formed, and the possible role of
non-native
interactions~\cite{Sulkowska:2009:Proc-Natl-Acad-Sci-U-S-A:19211785,Noel:2010:Proc-Natl-Acad-Sci-U-S-A:20702769,Wallin:2007:J-Mol-Biol:17368671,ourplos}.

To ease the major computational burden imposed by simulating the slow
process of spontaneous folding/knotting of these molecules, the
above-mentioned studies were performed using   $G\hat{o}$-type native-centric force
fields, in either coarse-grained (CG)
 or atomistic protein representations. The latter approach allowed for establishing the noteworthy
result that by promoting native interactions alone it is possible to
fold a natively-knotted
protein~\cite{Sulkowska:2009:Proc-Natl-Acad-Sci-U-S-A:19211785,Noel:2010:Proc-Natl-Acad-Sci-U-S-A:20702769}.
Non-native interactions have, however, been argued to be important for
enhancing the efficiency of the process, by significantly increasing the
accessibility of knotted configurations in the early folding
stages~\cite{Wallin:2007:J-Mol-Biol:17368671,ourplos}.

A natural test case for numerical studies of spontaneous knotting in polypeptide chains is represented by protein MJ0366, which is the shortest known knotted protein. The folding process of this 82-amino acid long protein appears to be governed by such a delicate interplay of amino acid stereochemical interactions that folding simulations employing  different levels of spatial resolution have been shown to  yield different knotting mechanisms. In particular, the seminal study of Noel and co-workers \cite{Noel:2010:Proc-Natl-Acad-Sci-U-S-A:20702769}, where the folding of MJ0366 was characterized using pure native-centric force-fields, has shown that in coarse-grained folding simulations, the knot could form at either terminus, while only the C-terminal is involved in knotting when the full atomistic detail is used.

The observed sensitivity of the MJ0366 folding process on structural details poses a further fundamental question:  to what extent is the knotting mechanism sensitive to details of the force field used in folding simulations?. Towards this goal, we here analyze an ensemble of about 30 successful atomistic folding trajectories for protein MJ0366, obtained by using a realistic force field, namely AMBER99ffSB~\cite{AMBER99sb} with implicit solvent.\\ To the best of our knowledge this represent the
  first instance where a realistic force-field is employed to follow the
  folding of initially unfolded, and unknotted conformations into a
  knotted native state.

To collect this sizeable number of productive
  trajectories in an affordable amount of computational time, we have
  used an advanced simulation technique known as the ``dominant reaction
  pathway'' (DRP) scheme. In other protein contexts, this method was
  shown to yield results consistent with standard extensive MD folding
  simulations, performed with the same atomistic force
  field~\cite{DRPFIP}.

We find that self-knotting of MJ0366 typically occurs at a late folding
stage, when about $90\%$ of the native contacts are established and
almost invariably involves a single dominant knotting mechanism. The
latter consisting of the threading of the C-terminus through an open
region created by an already formed $\beta$-sheet.  Based on
  various model calculations it is argued that the observed difference
  in knotting modes is strongly influenced by non-native interactions.

\section*{Results/Discussion}

\begin{figure*}
\includegraphics[width=0.4\textwidth]{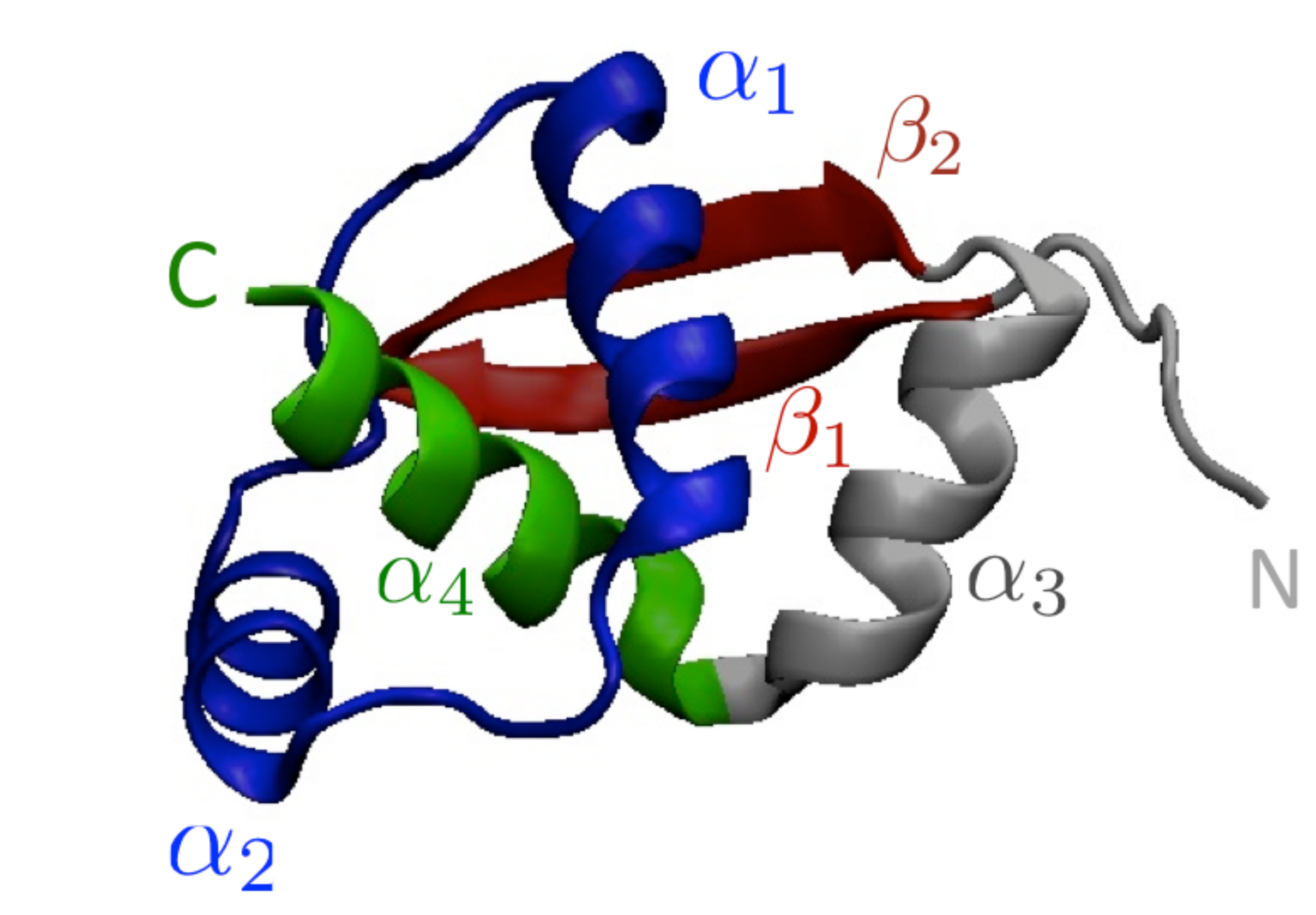}
\caption{{\bf Crystal structure of protein MJ0366, PDB code: 2EFV. This and other images were rendered with VMD\cite{Humphrey:1996:J-Mol-Graph:8744570}.}}
\label{fig1}
\end{figure*}

The monomeric unit of the natively-knotted MJ0366 protein consists of 82
amino acids and comprises four $\alpha$-helices and one $\beta$-sheet
resulting from the pairing of two antiparallel strands with a large
sequence separation ($\sim$30 amino acids), see Fig.~\ref{fig1}. The
C-terminal helix, $\alpha_4$, protrudes through a loop formed by the other two
$\alpha$-helices giving rise to a rather shallow trefoil-knot.

We report on the characterization of the folding process of MJ0366 by
means of advanced molecular dynamics techniques based on the AMBER99ffSB
atomistic force field~\cite{AMBER99sb} with implicit solvent. The
numerical strategy was articulated over several steps. 

Specifically, we first generated an ensemble of 100 denatured
configurations for protein MJ0366 by unfolding the crystal structure
using 100 ps of atomistic molecular dynamics (MD) simulations at high
temperature (1600 K) followed by 100 ps of thermalization at 300~K.

Next, the folding and knotting dynamics of MJ0366 was studied by
carrying out 40-50 folding attempts for each of the 100 denatured
configurations, for a total of about 4000 attempted folding
trajectories.  Simulating so many folding trajectories from an initially
unfolded state is presently beyond reach of standard MD simulations even
when run on dedicated supercomputers \cite{Shaw2010}.  To overcome these
difficulties  we resorted to the recent development of the DRP
  approach proposed in ref.~\cite{DRPFIP}. This combines a ratchet-and-pawl
  molecular dynamics algorithm \cite{rMD1,rMD} (rMD) with a statistical
  analysis based on scoring {\em a posteriori} the relative likelihood of each computed
  folding pathways \cite{DRP1,DRP2,DRP3,DRP4}. This method is described
  in detail in the next section, and has been recently used to
  investigate the folding of the WW domain FIP35~\cite{DRPFIP} yielding
  a very consistent folding mechanism with ms-long MD simulations in
  explicit solvent~\cite{Shaw2010}. 

The strength of the rMD scheme is that it allows for
efficiently generating an ensemble of trial folding pathways from a
given initial denatured state to the known native state, while keeping
at a minimum the external work applied to drive the system. In
fact, the system dynamics evolves in a completely unbiased way whenever
it leads to a higher similarity with the native state, i.e. a larger
number of formed native contacts. Conversely, a time-dependent external
force is introduced to discourage, though not completely prevent, a
decrease of the native similarity. The biased rMD evolution
  promotes only the overall geometrical similarity with the native state
  and does not reward specific concerted backbone movements that could
  lead to knotting. As a matter of fact, knot formation was observed
  only for a small fraction of the thousands of attempted rMD
  trajectories, namely 66 of them, covering 31 distinct
  initially-unfolded states. In all cases, the knotting event
  corresponded to the formation of the native trefoil knot, thus
  indicating that incorrect knot formation is not a major source of
  kinetic trapping for MJ0366.

 In the DRP approach, only one productive pathway per initial
  condition was retained, namely the one with the highest statistical
  weight. This weight, corresponds to the probability that each trial
  trajectory is generated by an overdamped Langevin dynamics. Notice
  that, because the weights are calculated with reference to an {\em
    unbiased} stochastic dynamics, the DRP selection criterion lessens
  {\em a posteriori} the rMD steering effects.\\

{\bf Trajectories analysis.} The selected 31 trajectories were analyzed by monitoring the evolution
of several geometrical and topological parameters during the folding
process.

\begin{figure*}
\includegraphics[width=0.4\textwidth]{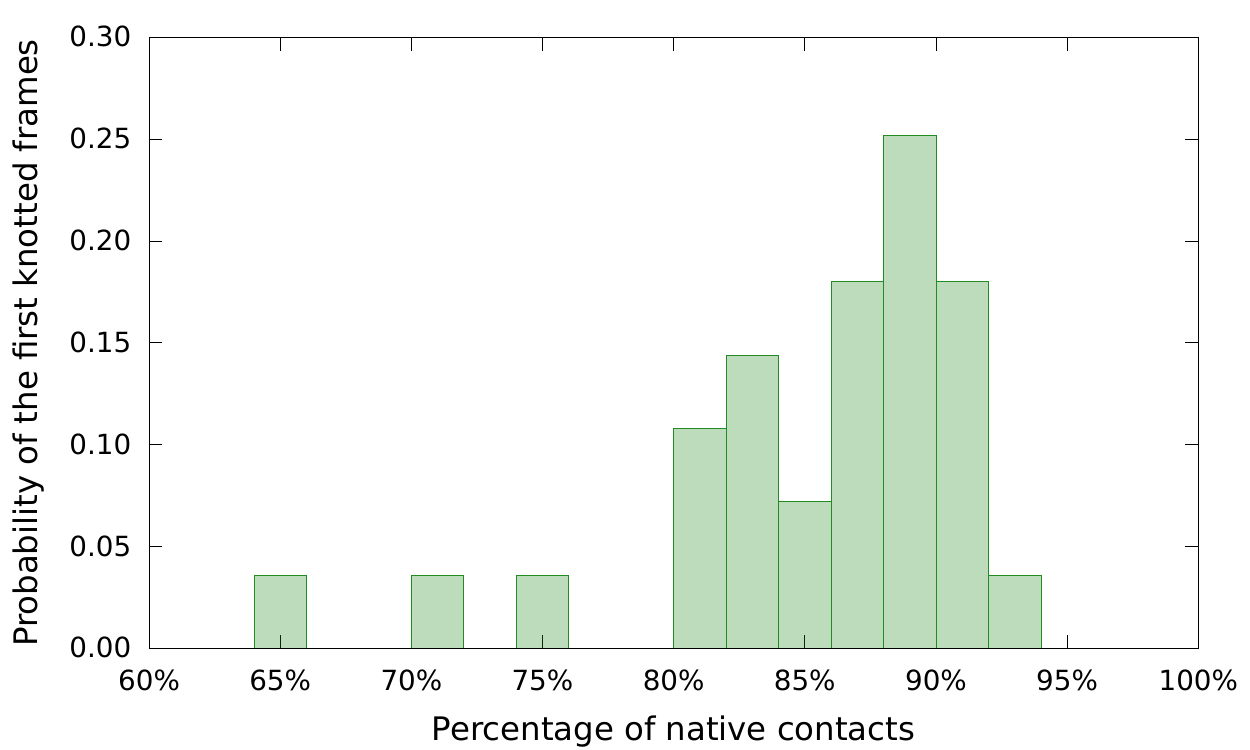}
\caption{{\bf Distribution of the percentage of formed native contacts at the time of the first knotting event for the 26 DRP trajectories of MJ0366.}}
\label{fig2}
\end{figure*}

  As a first step we identified the folding stage at which the backbone
  self-ties into knot. Accordingly, for each trajectory we calculated
  the percentage of native contacts (overlap) that are formed when the
  first knotting event occurs.  The distribution of these overlaps for
  the considered trajectories is shown in Fig.~\ref{fig2}. The
  distribution is peaked at about 90\% overlap. This indicates that the
  knot is typically formed at a rather late stage of the folding
  process.

Next, to characterize the diversity of the folding pathways and the
implications for the knotting mechanism,
we computed the average ``path similarity parameter'', $s$. As explained in the
Materials and Methods section, this quantity measures the consistency of the temporal
succession in which the native contacts are formed in two given
pathways.  The $s$ parameter takes on values ranging from 0, for no
similarity, to 1 when all native contacts form with exactly the same
succession in the two trajectories. We emphasize that $s$ depends only
on the time order of native contact formation events (and not their exact
timing). 

\begin{figure*}
\includegraphics[width=0.4\textwidth]{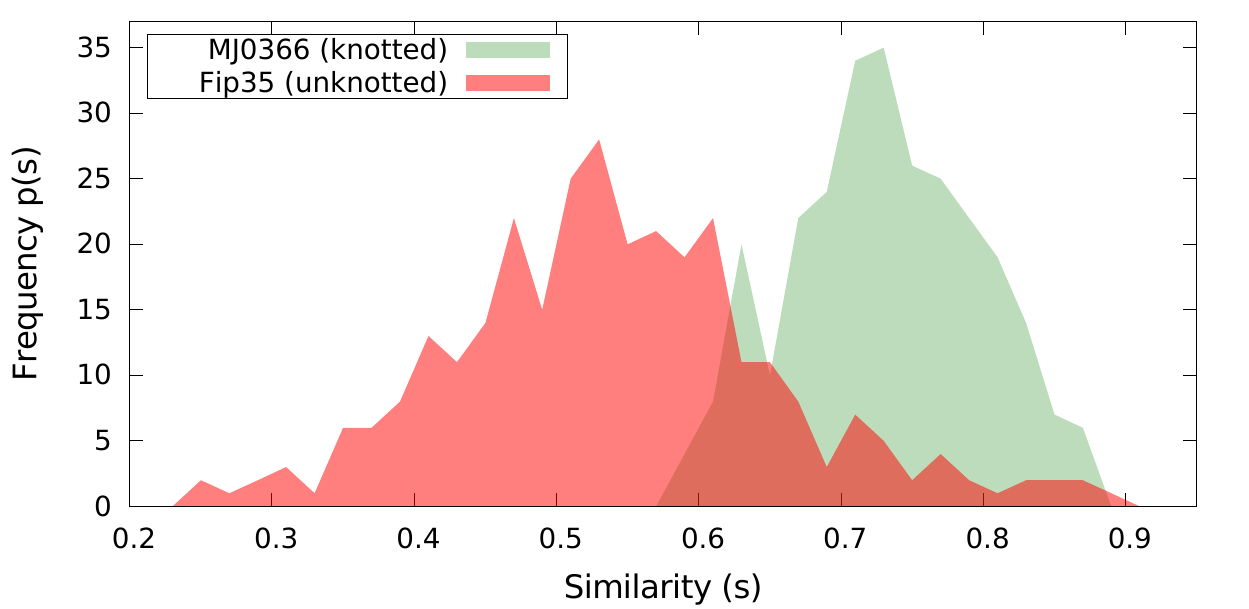}
\caption{
{\bf Distributions of the path similarity parameter $s$, see Eq. \ref{pathdiversity} for DRP trajectories. The green distribution pertains to the 26 DRP trajectories of the knotted protein MJ0366. For comparative purposes, the red curve shows the $s$ distribution of DRP trajectories of the uknotted WW domain FIP35\cite{DRPFIP}.}}
\label{fig3}
\end{figure*}

To have a robust indication of the degree of heterogeneity of the
selected trajectories, we computed the distribution of $s$ over all
possible pairs of trajectories, see Fig.~\ref{fig3}. As a term of
reference, the same Figure shows the $s$ distribution computed over
previously-studied folding trajectories of the unknotted WW domain
FIP35~\cite{DRPFIP}. It is seen that the distribution of MJ0366 is
narrower and shifted towards significantly higher values of $s$ than for
the unknotted protein. Indeed the former has a peak at $s\sim 0.75$
while the latter has it at $s\sim 0.5$. This relatively low value of $s$
and the distribution broadness is typical of folding processes that
proceed by multiple pathways~\cite{DRPFIP,Krivov2011}, as FIP35 is known
to do. The different characteristics of the $s$ distribution for MJ0366
therefore strongly suggest the existence of one dominant folding pathway
for MJ0366.

 We accordingly sought to analyze in detail the folding process to
 verify that knotting occurs via one dominant mechanism and characterize
 it.

In this regard a valuable term of reference is given by the earlier
  study of Noel {\em et al.}
  \cite{Noel:2010:Proc-Natl-Acad-Sci-U-S-A:20702769} where the folding
  thermodynamics of MJ0366 was systematically characterised with both
  atomistic and coarse-grained native-centric models.  When the
  atomistic model was employed, it was seen that knotting preferentially
  occurred via slipknotting. Specifically, in most of the productive
  trajectories obtained at the folding temperature of the
  structure-based model, the C-terminal attained a hairpin-bent
  conformation and established the knot by threading the open region
  involving residues 17-54.  The slipknotting
  mechanism was found to occur more frequently than that of other knotting modes, such
  as the threading of the open region by a non-bent C-terminus, or knot
  formation at the N terminus. Interestingly, the coarse-grained
    native-centric model was more prone to unproductive
    kinetic traps and displayed significant heterogeneity for
    knotting mechanisms too. These aspects indicated that the realistic
    treatment of protein structural detailed significantly helped reduce
    the impact of unproductive routes in the folding process
    \cite{Noel:2010:Proc-Natl-Acad-Sci-U-S-A:20702769}.

Here, by addressing the same protein folding process with a
  realistic, non native-centric force-field, it is possible to
  examine to what extent various aspects of the knotting process are
  sensitive to the treatment of inter-atomic interactions.

As a first step of the analysis, we profiled the folding trajectories
along two relevant order parameters: the root mean square distance
(RMSD) to the native structure and the RMSD to the native $\beta$-sheet.
The first collective variable monitors the overall progress towards the
native geometry. The second one, instead, carries information about one
of the expected entropic bottlenecks of the folding process, namely the
formation of  the native antiparallel $\beta$-sheet which involves
amino acid pairs with a sequence separation as large as 38.

Since in the native MJ0366 structure the C-terminal helix protrudes
through the region intervening between the two paired $\beta$-strands,
monitoring the formation of the $\beta$-sheet is relevant to understand
whether the sheet is formed before or after the knot.

\begin{figure*}
\includegraphics[width=0.8\textwidth]{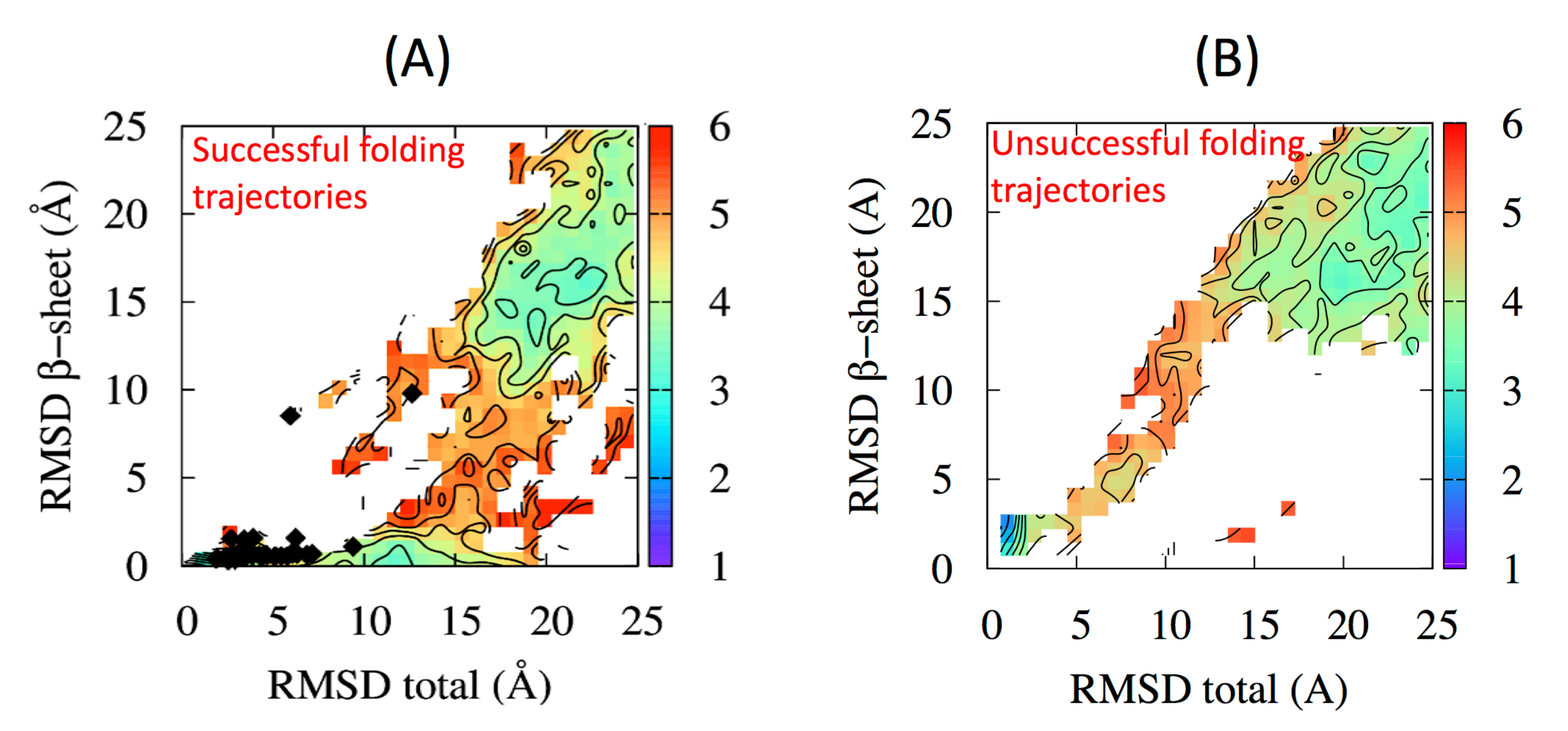}
\caption{
{\bf Atomistic DRP folding pathways, projected on the plane selected by the total RMSD to native and by the RMSD to native of the $\beta$-sheet. Panels (A) and (B) refer respectively to successful and unsuccessful folding trajectories. The diamonds in panel (A) mark the collective coordinates at the time of knot formation. The scale on the left corresponds to the logarithm of the number of times a given spot is visited by the DRP trajectories, in analogy with free-energy landscape plots. }}
\label{fig4}
\end{figure*}

The results shown in the left panel of Fig.~\ref{fig4} indicate that the $\beta$-sheet is
fully formed rather early, when the total RMSD to native of the chain is
about 15 \AA. At this stage the fraction of formed native contacts is
about 40-50 $\%$. The self-tying of the molecule into a trefoil knot
typically occurs after the formation of the $\beta$-sheet. This is
evident from the placement of the diamond symbols in Fig.~\ref{fig4}
which mark the first occurrence of knots for each of the 31
trajectories. It is seen that all first-knotting events occur when the
$\beta$-sheet is fully formed, with only two exceptions that will be discussed later.

\begin{figure*}
\includegraphics[width=0.8\textwidth]{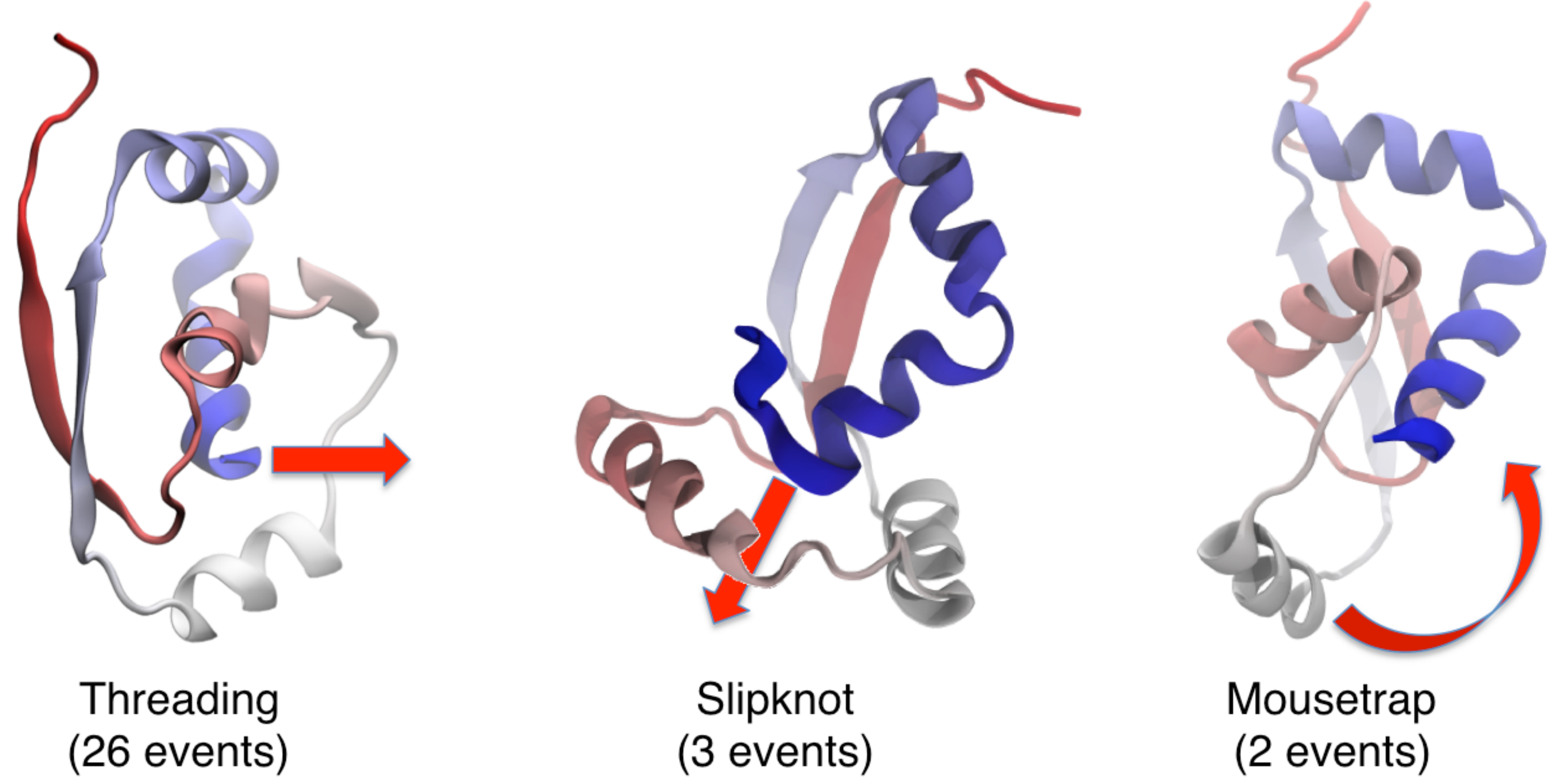}
\caption{{\bf The three different types of knotting mechanisms observed in our atomistic DRP simulations.}}
\label{fig5}
\end{figure*}

 The detailed inspection of the trajectories indicates that the
  knotting process almost invariably occurs through the so-called
  "threading" mechanism, where the $C$-terminal $\alpha$-helix (residues
  74-87) directly enters, without bending, the open region between amino
  acids 17-54 involving helices $\alpha_1$ and $\alpha_2$ and the intervening loop, see
the sketch in the left panel of Fig.~\ref{fig5}.  In this case, the
threaded region and the $\beta$-sheet
(respectively shown in blue and red  in Fig.~\ref{fig1}) establish a tertiary contact before the terminal helix
penetrates into the open region in between the helices $\alpha_1$ and
$\alpha_2$ (see left panel in Fig.~\ref{fig5}). This mechanism accounts
for as many as 26 of the 31 rMD trajectories.

 In three other cases, the folding was found to occur through the
  so-called "slipknot"
  mechanism\cite{Noel:2010:Proc-Natl-Acad-Sci-U-S-A:20702769}
  (i.e. where the open region is entered by the backward-bent
  C-terminus). In all three instances the $C$ terminus entered the
loose $\alpha_1$--$\alpha_2$ region producing a shallow slipknotted
trefoil, as shown in the central panel of Fig.~\ref{fig5}.

Finally, in two further cases we observed another knotting mechanism
which involves a concerted backbone movement that had not been
previously reported for MJ0366. Specifically, in two trajectories when
the $\beta$-sheet and the terminal $\alpha$-helix are already formed and
juxtaposed in an unknotted configuration the loop performs a
``mousetrap-like'' movement establishing the native knotted
topology. This movement, which bears some analogies with the
suggested knotting mechanism for an unrelated protein with a non-trefoil
topology~\cite{virnauplos}, is schematically represented in the right
panel Fig.~\ref{fig5}. The mousetrap knotting events correspond to the
two outlying diamonds reported in Fig.~\ref{fig4}, with collective
coordinates (6\AA,~8\AA) and (12\AA,~10\AA). 

Videos obtained from the atomistic DRP trajectories which illustrate the
three observed knotting mechanisms are included in the on-line SI.

\begin{figure*}
\includegraphics[width=1.0\textwidth]{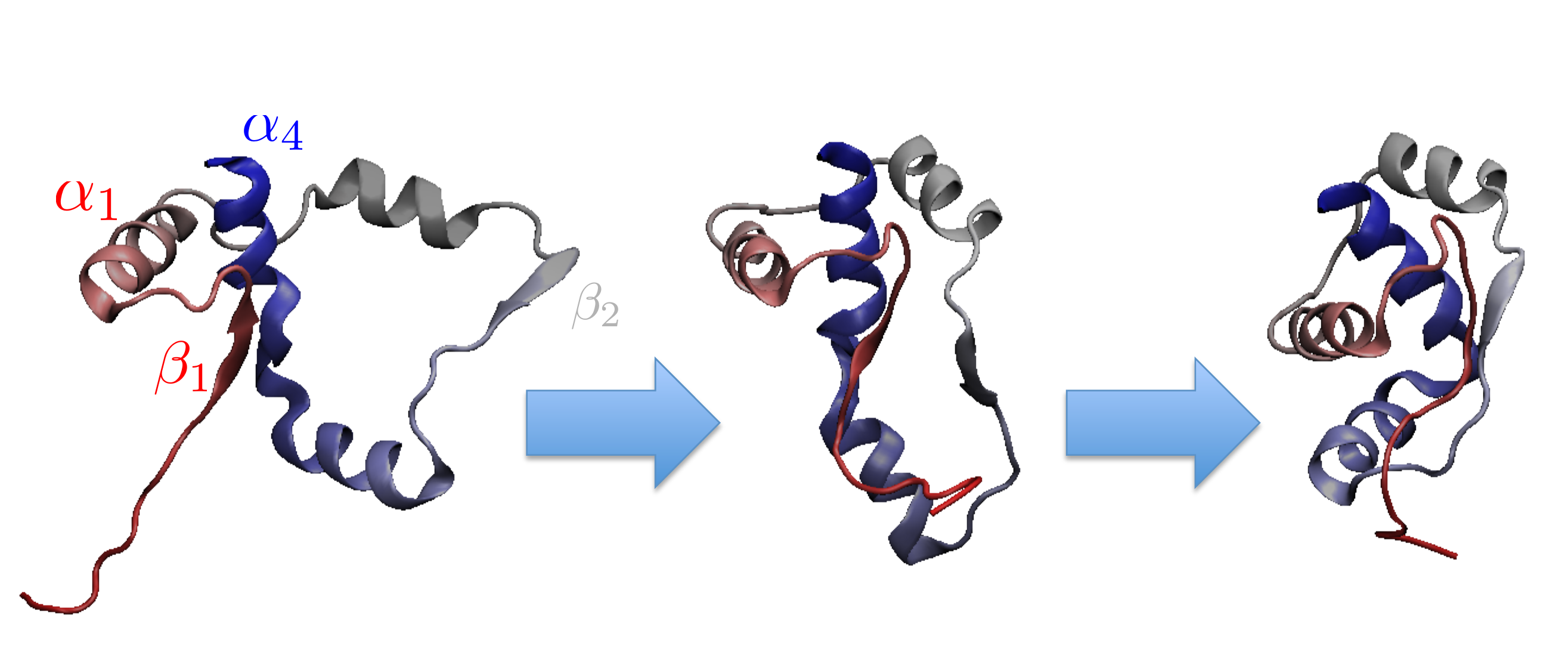} 
\caption{{ \bf Exposure to the solvent of  polar, non-polar and charged residues along the folding trajectories pertaining to three different knotting mechanisms, plotted as a function of the RMSD to the native structure. The number of amino acids exposed to the solvent was computed using the VMD utility\cite{Humphrey:1996:J-Mol-Graph:8744570}. The dashed and dot-dashed lines represent folding events with mousetrap and slipknotting mechanism, respectively. The points are the average  over the 26 DRP trajectories with a threading knotting mechanism, and the error bar denote the corresponding standard deviation. Left panel: evolution of non-polar residues, Right panel: evolution of polar and charged residues. }}
\label{fig6}
\end{figure*}

 It is important to notice that the trajectories associated to the
  various knotting modes do not present significant quantitative
  differences regarding the overall solvent accessibility of polar and
  non-polar residues during the folding process.  This point is
  illustrated in Fig.~\ref{fig6} where the number of exposed
  hydrophobic and hydrophilic residues are profiled versus the RMSD to
  the native state. The consistency of the various profiles provides a
  quantitative basis for expecting that the relative weight of the
  knotting mechanisms should not depend critically on the specific model
  adopted to describe the solvent-induced interactions.\\

{\bf Order of contact formation and knotting.} To understand how the interplay of amino acid interaction captured
  by the realistic force field favours knotting by threading, we
  have carried out a comparative analysis of the reaction mechanism in
  successful and unsuccessful folding trajectories. Specifically, the
  productive, successful set consisted of the 26 trajectories displaying
  the dominant (threading) knotting mechanism. The non-productive one
  included an equal number of trajectories that reached an {\em
    unknotted} configuration and nevertheless had a good native
  similarity (namely an RMSD to the crystal structure less than 5\AA). 

The projection of the unsuccessful trajectories along the two  collective
  coordinates considered before is shown in Fig.\ref{fig4}B. The
  qualitative difference with respect to the analogous plot for the
  successful ones (panel A) is striking. In particular, it is seen that
  in successful trajectories the formation of the sheet involving
  strands $\beta_1$ and $\beta_2$ occurs rather early on and prior to
  the establishment of the overall tertiary organization of MJ0366. In
  fact, the total RMSD to native decreases appreciably only after the
  $\beta$-sheet is established. By converse, for unsuccessful
  trajectories, this hierarchy of contacts formation is not observed,
  and the $\beta$-sheet formation proceeds in parallel with the
  acquiring of the overall native structure. One therefore concludes
  that the early formation of the $\beta$-sheet provides the most
  appropriate conditions for knotting by leaving the region delimited by
  the $\beta$ sheet accessible to threading events.

\begin{figure*}
\includegraphics[width=.7\textwidth]{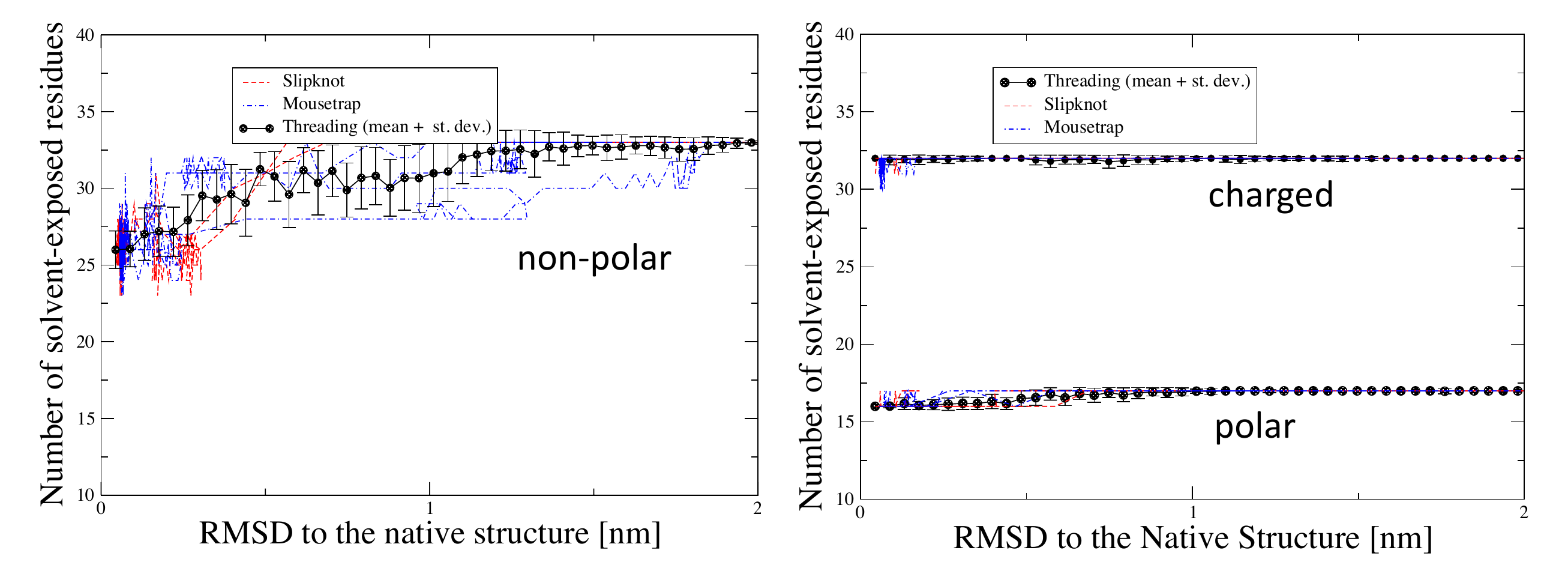} 
\caption{{\bf Typical example of unsuccessful trajectory. The late formation of the $\beta$-sheet traps the $N$ terminus on the ``wrong'' side of the $\beta_2-\alpha_3$ loop and prevents attaining the (native) knotted topology.}}
\label{fig7}
\end{figure*}

This conclusion is supported by the detailed inspection of the
  unsuccessful trajectories, which are exemplified in the sequence of
  snapshots shown in Fig.\ref{fig7}. As it is visible in this figure,
  the C-terminal helix threads the correct region between strands
  $\beta_1$ and $\beta_2$ prior to the formation of the
  $\beta$-sheet. When the latter is finally establishes, the
  $N$-terminal segment remains trapped on the wrong side of the loop
  bridging $\beta_2$ and $\alpha_3$ and, for steric reasons cannot go
  past it and attain the native knotted topology.

The relevance of this mechanism for misfolding is
  highlighted by the fact that all unsuccessful trajectories displayed a
  late formation of the $\beta$-sheet. We emphasize again that,
  according to our simulations, the correct knotting of the chain is not
  promoted by the formation specific contacts which fail to form in
  misfolding events. Rather, for the chain to acquire the native
  topology, it is essential that the native contacts form in the correct
  order.\\

{\bf Further insight from coarse-grained models.} The fact that the observed dominant knotting mode differs from the
  one reported previously using pure native-centric force fields
  suggests that non-native interactions could be relevant for
  favouring the correct succession of contacts leading to self-knotting
  (or avoiding unproductive ones).  This possibility is particularly
  interesting in connection with the ongoing discussion about the role
  that non-native interactions can have in aiding the knotting process
  even during the early
  folding~\cite{Wallin:2007:J-Mol-Biol:17368671,ourplos}.

To investigate this aspect we generated several folding
  trajectories for MJ0366 using simplified models where the effect of
  non-native interactions could be easily turned on or
  off. Specifically, we considered two different coarse-grained models:
  one with only native-centric interactions and the other additionally
  incorporating non-native interactions. The latter included
  quasi-chemical and screened electrostatic pairwise interactions, as in
  the recent study of the early folding stages of a trefoil-knotted
  carbamoyltransferases~\cite{ourplos}.

The folding process presents major differences in the two
  models. First, they differ significantly in terms of knotting
  probability. Specifically, for each model we considered an extensive
  set of 10,000 uncorrelated configurations, equilibrated at the nominal
  temperature of 300 K. In the native-only case, 12\% of the sampled
  configurations were knotted, while this number had a sixfold
  increased, to 75\%, in presence of non-native interactions. This
  result aptly complements the atomistic DRP simulations, for
  highlighting the role of non-native interactions in aiding the
  formation of the native knotted topology of MJ0366.

Secondly, productive trajectories follow different dominant
  mechanisms in the two models. In fact, when the pure native-centric
  model is used, 8 out of the 10 trajectories involved the slipknotting
  mechanism, while the threading one was observed in all trajectories
  (10 out of 10) with the additional non-native interactions. The latter
  result, which is in full accord with the atomistic DRP simulations,
  reinforces the concept that non-native interaction can promote the
  correct order of contact formation required for self-knotting.

\begin{figure*}
\includegraphics[width=0.8\textwidth]{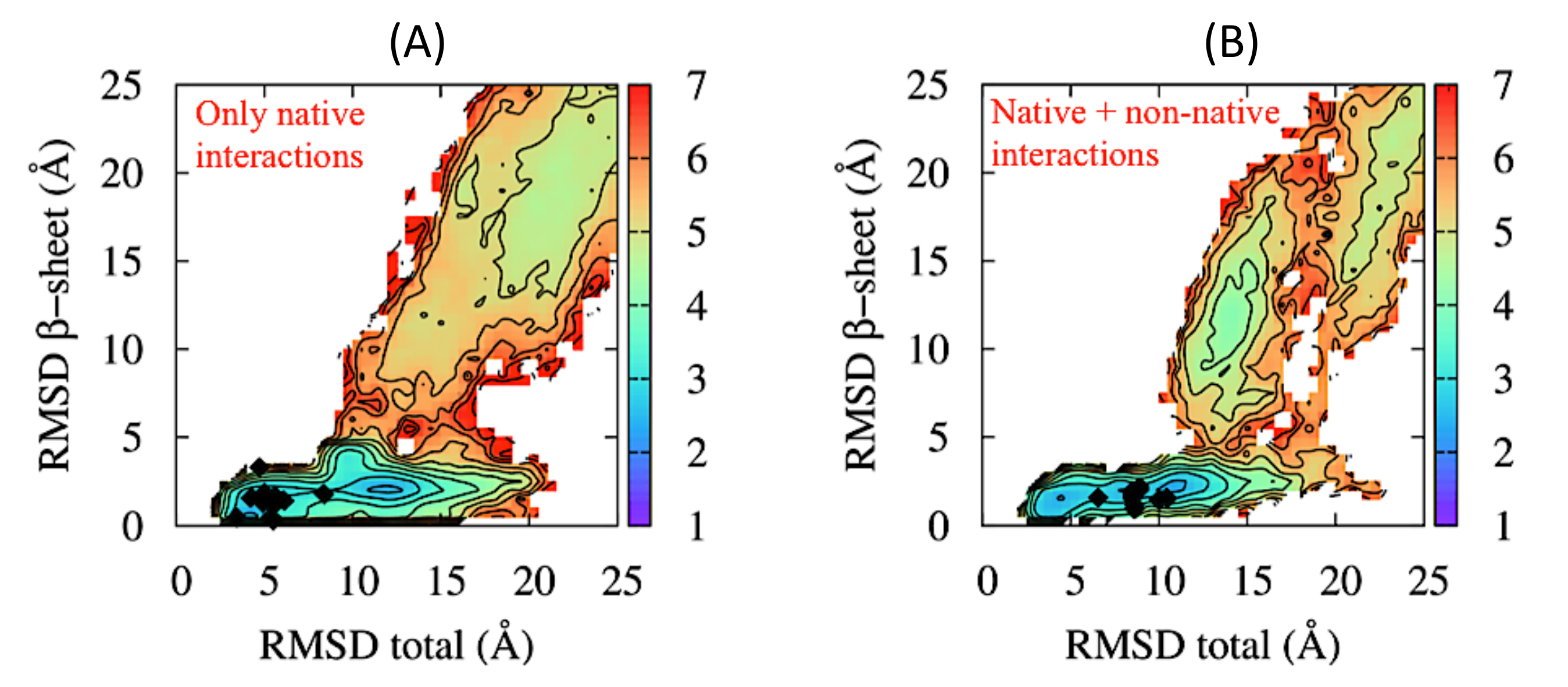} 
\caption{{ \bf Folding pathways obtained from coarse-grained Monte Carlo
    simulations with local crankshaft moves which mimic the chain
    dynamics, projected on the plane selected by the total RMSD to
    native and by the RMSD to native of the $\beta$-sheet.  Panel (A) refers to
    the model with only native interactions, while panel (B) refers to
    the model with both native and non-native interactions. The diamonds
    denote the values of the collective coordinates at the time of knot
    formation. The scale on the left is the logarithm of the number of
    times the point is visited by folding trajectories, in analogy with
    free-energy landscape plots. }}
\label{fig8}
\end{figure*}

This point is further supported by the inspection of
  the density plots in Fig.~\ref{fig8}.  In fact, non-native
  interactions are more clearly associated to the early formation of the
  $\beta$-sheet than for the native-only case. Furthermore, the path
  outlined in panel B bears more analogies than the one in panel a with
  the density plot of Fig. \ref{fig4}A, which captured the successful
  folding events obtained from atomistic DRP simulations. Indeed, in the
  simplified model, the early formation of the $\beta-$sheet is promoted
  by the fact that the non-native quasi-chemical interaction generates
  an overall attractive interaction between the residues in $\beta_1$
  and those in $\beta_2$. Consistently with the misfolding events
  discussed previously, one can therefore argue that the weaker drive of
  the native-centric model to form early on the $\beta$-sheet, is also
  responsible for its lower knotting propensity.

Based on these results, we can argue that mutations in
  the $\beta-$sheet regions with residues characterized by a weaker
  effective attraction, would delay the formation of the $\beta$-hairpin
  in the folding process and would make the chain more prone to reach
  the unknotted mis-folded state. This prediction may be verified
  experimentally. \\

{\bf Concluding remarks.} In conclusion, the DRP simulations presented here
  provided, to the best of our knowledge, the first systematic attempt
  to characterize the folding process of a natively-knotted protein,
  MJ0366, using a realistic atomistic force field. MJ0366 knotting is
  observed to occur via threading at the C-terminal. The comparison of
  productive and unproductive trajectories (which respectively end up in
  natively-knotted and unknotted states) further indicates that knotting
  is aided by the early formation of the native $\beta$-sheet.  By
  comparing the MJ0366 knotting propensity and mechanisms in simplified
  folding models it is argued that non-native interactions are important
  for aiding knotting by promoting the correct order of contact formation.

 While there is no {\em a priori} reason to expect that non-native
  interactions are crucial for guiding the folding process of knotted
  proteins in general, it is interesting to notice that their important
  role has been previously suggested for another trefoil-knotted
  protein carbamoyltransferases~\cite{Wallin:2007:J-Mol-Biol:17368671,ourplos}. In our
  view, it would be most interesting to further examine this effect, in
  future studies on MJ0366 or other proteins either through experiments
  (e.g. involving mutagenesis) or with more extensive simulations,
  possibly involving explicit solvent treatment or unbiased dynamics.

\section*{Materials and Methods}

{\bf The rMD and DRP algorithms}: In order to generate an ensemble of trial trajectories connecting a given initial configuration to the native state we used the following variant of the rMD algorithm. At each integration step, we evaluated a collective coordinate (CC) which measures the distance of between the instantaneous contact map and the 
native contact map: 
\be
 z[{\bf X}(t)] \equiv \sum_{i<j}^{N} [ C_{ij}[{\bf X}(t)] - C_{ij}({\bf X}^{\text{native}}) ]^2, 
\ee
with $j-i~>~35$ with a distance cutoff of $12.3$~\AA.
In this equation, ${\bf X}$ is the 3N-dimensional vector in configuration space, and $C_{ij}[{\bf X}(t)]$ and $C_{ij}({\bf X}^{\text{native}})$ are the instantaneous and native contact map, respectively. 
The entries of the contact map C$_{ij}(X)$ are chosen to  interpolate smoothly between 0 and 1, depending on the relative distance of the 
atoms $i$ and $j$:  
\be
C_{ij}({\bf X}) = \{ 1-(r_{ij}/ r_0 )^6\}/\{ 1-( r_{ij}/ r_0 )^{10}\},
 \ee
where r${_0}$=7.5~\AA~ is a fixed reference distance. 

In the rMD algorithm, no bias is applied to the chain when it spontaneously diffuses towards the bottom of the folding funnel, i.e. any time the value of the CC at time $t+\Delta t$ is smaller than the minimum value so far. 
On the other hand, fluctuations which would drive the contact map further from the native one (hence increasing $z$)  are hindered by introducing a biasing force, defined by the time-dependent potential 
\be
  V_R({\bf X}, t) = \left\{
    \begin{array}{lr r}
      \frac{k_R}{2} ( z[{\bf X} (t)] - z_m(t) )^2, &\text{for}&\quad z[{\bf X}(t)] > z_m(t)\\
      0, &\text{for}&\quad z[{\bf X}(t)] \leq z_m(t).\\
    \end{array}
  \right.\vspace{2mm}
  \label{VR}
\ee
In these equations, $k_R=0.02$ kcal/mol is the so-called ratchet constant and  $z_m(t)$ is the minimum value assumed by the collective variable $z$ along the rMD trajectory, up to  time~$t$. 

In the original formulation of the rMD algorithm \cite{rMD}, the variable  $z_m(t)$ is updated
only when the system visits a configuration with $z[{\bf X}(t+\delta t)]< z_m(t)$.  With this choice, $z_m$ monotonically 
decreases during the course of the simulation.  In this work, we choose to significantly  weaken the effect of the bias by allowing the system to backtrack along the direction defined by the CC. This is done by occasionally updating  $z_m$
also when it increases, according to a Metropolis accept/reject criterium. Namely, $z_m$ is updated to  $z^{'}_m=z[{\bf X}(t+\delta t)]>z_m$ 
if $\exp\{-\beta_{r_{MD}}[(z'_m-z_m)-2 (z'_m-z_m)^3]\}< \eta$, where $\eta\in [0,1]$ is a random number sampled from a uniform distribution and $\beta_{r_{MD}}=3 \times 10^{-2}$ is an artificial "inverse thermal energy". 
This modification of the original rMD algorithm  is required to escape from kinetic traps. 
Without it the folding efficiency to the correct topologically non-trivial native state is strongly suppressed. Each trial trajectory consisted of $1.8 \times 10^5$ steps of rMD with a nominal integration time step of $\Delta t=1$~fs.

The DRP algorithm is used to identify the most probable path in each set of trial rMD trajectories sharing the same boundary conditions. This is done by evaluating the relative probability for each path $X$ to be realized in the unbiased over-damped Langevin dynamics: 
\be
\text{Prob}[{\bf X}] = e^{-\sum_{i=1}^{N_t} \sum_{k=1}^N
\frac{1}{4 D_k \Delta t} \cdot \left({\bf x}^k(i+1)-{\bf x}^k(i) + \frac{\Delta t ~D_k}{k_B T} \nabla_k U[i] \right)^2}\ .
\ee
In this equation, the index $i=1,\ldots, N_t$ runs over the different time-step in the  trajectory, the index $k=1,\ldots, N$ runs over the atoms in the protein, $k_B$ is the Boltzmann's constant  and $D_k$ is the diffusion coefficient of the $k$-th atom. 

{\bf Atomistic force field}: In both rMD and standard high-temperature MD simulations we used the AMBER~ff99SB force
field~\cite{AMBER99sb} in implicit solvent, within the 
Generalized Born formalism implemented in GROMACS 4.5.2 \cite{GRO4}. In such an approach, the Born radii are calculated according to the Onufriev-Bashford-Case algorithm \cite{OBC}.  The hydrophobic tendency of non-polar residues is taken into account through an interaction term proportional to the solvent-accessible-surface-area (SASA). The solvent-exposed surface of the different atoms is calculated from the Born-radii, according to the approximation developed by Schaefer, Bartels and Karplus in \cite{SASA}.

{\bf Alternative simplified force fields}: The CG folding simulations were based on the model developed in Ref.s~\cite{Kim&Hummer,reactioncoord2}. In this approach,  amino acids are represented by spherical beads centered at the $C_\alpha$ positions. The non-bonded part of the potential energy contains both native and non-native interactions.  The former are the same used in the G$\overline{o}$-type model of Ref.~\cite{Karanikolas&Brooks}, while the latter consist of a quasi-chemical potential,
  which accounts for the  statistical propensity of different amino-acids to form contact and of a Debye-screened electrostatic term. A detailed description of the force field of this model can be found in Ref.~\cite{ourplos}.   In our previous work, we have shown that such non-native interactions are able to strongly promote the knot formation in natively knotted polypeptides~\cite{ourplos}.
Folding simulations for protein MJ0366 in this CG model were performed using a MC algorithm described in detail in Ref. \cite{ourplos}. 
This type of crankshaft-based MC algorithm is commonly
employed in polymer physics~\cite{fast_MC_algorithms} to study 
dynamic properties, since it is was shown that they mimic the intrinsic dynamics 
of a polymer in solution~\cite{binder_crankshaft} at a much lower computational 
cost than standard MD simulations~\cite{mc_for_comformational_sampling}. 
The folding dynamics of CG model with native and non-native
interactions was simulated by generating 200 MC trajectories,
while the dynamics of the model with only native interactions
was investigated by generating 500 MC trajectories. For both CG models, trajectories consist of 1.5$\times$10$^8$ attempted MC moves,
corresponding to 1.5$\times$10$^4$ saved frames. MC moves that we have employed were
the local crank-shaft and Cartesian moves, whose boldness was chosen
such that the acceptance rate was nearly constant and approximately equal to 50$\%$. In both cases, we have collected a total of 10 folding transitions, leading
to native configurations with the correct knotted topology. In order to compute the frequency of knotted configurations at thermal equilibrium we performed MC simulations which combine local moves and global pivot moves. 

{\bf Knot detection}: 
 The conformations visited during the MC dynamics were analysed to establish their global and local knotted state.
The global topological state was established and assigned by computing the Alexander determinants after suitable closure of the whole protein chain into a ring.
 For each configuration, this entailed 100 alternative closures where each terminus is prolonged far out of the protein along a stochastically chosen direction, and the end of the prolonged segments are closed by an arc ``at infinity'' (i.e. not intersecting the protein). As in ref.~\cite{ourplos}, to avoid considering back-turning closures, stochastic exit directions are picked uniformly among those which  form an angle of more than 90$^o$ with the oriented segment going from each terminus to the C$_\alpha$ at a sequence distance of 10. 
 If the majority of the 100 stochastic closures return non-trivial Alexander determinants, then the whole conformation can be considered as  globally knotted.
Because protein knotting can occur through slipknot formation \cite{Sulkowska:2012:Proc-Natl-Acad-Sci-U-S-A:22891304} , the global topology investigation
was complemented by a local one. In fact, a slipknot can be detecting by identifying a non-trivially knotted portion of a chain that has a different global topology, in our case the unknotted one.
To this purpose, we repeated the above-mentioned statistical closure scheme for all possible subportions of length 20, 30, 40, ... of the protein chain so to identify the smallest knotted, or pseudo-knotted, chain portion\cite{PTPS.191.192,Tubiana2011}.

{\bf Path similarity}: 
To quantitatively measure the folding pathways diversity we implemented
the analysis described in Camilloni et al. \cite{rMD}, that will be
shortly summarized in the following. A folding mechanism is here considered
to be a specific sequence of native contacts formation. Hence, for
each path we measured the time of formation of each native contact,
as the frame of the trajectory where the contact is first formed.
Given $t_{ik}$ as the time of formation of the $i^{th}$ native contact
in the $k^{th}$ trajectory, we computed for each path $k$ the matrix
$M_{ij}\left(k\right)$ defined as:
\begin{eqnarray}
M_{ij}\left(k\right)=\begin{cases}
1 & t_{ik}<t_{jk}\\
0 & t_{ik}>t_{jk}\\
\frac{1}{2} & t_{ik}=t_{jk}
\end{cases}
\end{eqnarray}
containing all the information regarding the folding mechanism as
defined above. For each pair of pathways $k,k'$ it is possible to
compute the similarity $s$ defined as
\be
s\left(k,k'\right)=\frac{1}{N_{c}\left(N_{c}-1\right)}\sum_{k<k'}\delta\left(M_{ij}\left(k\right)-M_{ij}\left(k'\right)\right),
\ee
$N_{c}$ being the total number of native contacts. The similarity
ranges from $0$ for completely different mechanisms, to $1$ for completely identical mechanisms.
 Finally, we consider the 
distribution 
\begin{equation}\label{pathdiversity}
p\left(s\right)=\sum_{k<k'}\delta\left(s-s\left(k,k'\right)\right)
\end{equation}
of the similarity parameter, evaluated from all pairs of  the folding pathways.

\section*{Acknowledgments}
The DRP approach was developed in collaboration with H. Orland, F. Pederiva and M. Sega. SaB, T\v{S}, RC and PF are members of the Interdisciplinary Laboratory for Computational Science (LISC), a joint venture of Trento University and the Bruno Kessler Foundation. 
Simulations were performed on the AURORA supercomputer located at the LISC (Trento). 

\newpage

\end{document}